\documentclass{article} 
\usepackage[preprint]{colm2026_conference}

\usepackage{microtype}
\usepackage{hyperref}
\usepackage{url}
\usepackage{booktabs}
\usepackage{multirow}
\usepackage[dvipsnames]{xcolor}
\usepackage{graphicx}      
\usepackage{adjustbox}
\usepackage{defs}

\usepackage{subcaption}

\usepackage{listings}
\usepackage{xcolor}
\usepackage{subcaption}

\definecolor{toolbg}{RGB}{247,248,250}
\definecolor{toolframe}{RGB}{120,130,145}
\definecolor{tooltext}{RGB}{35,39,47}
\definecolor{toolaccent}{RGB}{25,90,160}
\definecolor{toolstring}{RGB}{180,85,35}
\definecolor{toolcomment}{RGB}{90,120,90}

\lstdefinestyle{tooloutput}{
    basicstyle=\ttfamily\small\color{tooltext},
    backgroundcolor=\color{toolbg},
    frame=single,
    rulecolor=\color{toolframe},
    framerule=1.0pt,              
    framesep=6pt,
    breaklines=true,
    breakatwhitespace=false,
    columns=fullflexible,
    keepspaces=true,
    showstringspaces=false,
    upquote=true,
    tabsize=2,
    xleftmargin=0.4em,
    xrightmargin=0.4em,
    aboveskip=0.6em,
    belowskip=0.6em,
    postbreak=\mbox{\textcolor{toolaccent}{$\hookrightarrow$}\space},
    keywordstyle=\color{toolaccent}\bfseries,
    stringstyle=\color{toolstring},
    commentstyle=\color{toolcomment}\itshape
}

\lstdefinestyle{promptstyle}{
    basicstyle=\ttfamily\small,
    breaklines=true,
    breakatwhitespace=false,
    frame=single,
    xleftmargin=0.5cm,
    xrightmargin=0.5cm,
    backgroundcolor=\color{gray!10},
    showstringspaces=false,
    columns=flexible,
    keepspaces=true,
    escapeinside={(*@}{@*)},
}

\lstdefinestyle{cppcode}{
    language=C++,
    basicstyle=\small\ttfamily,
    keywordstyle=\color{blue}\bfseries,
    commentstyle=\color{gray}\itshape,
    stringstyle=\color{red},
    numbers=none,
    numberstyle=\tiny\color{gray},
    stepnumber=1,
    numbersep=5pt,
    backgroundcolor=\color{white},
    showspaces=false,
    showstringspaces=false,
    showtabs=false,
    frame=single,
    tabsize=4,
    captionpos=b,
    breaklines=true,
    breakatwhitespace=false,
    escapeinside={\%*}{*)},
    xleftmargin=2em,
    xrightmargin=2em,
    aboveskip=1em,
    belowskip=1em
}

\usepackage{amsmath}
\usepackage{amssymb}
\usepackage{mathtools}
\usepackage{amsthm}
\usepackage{listings}
\lstdefinestyle{promptstyle}{
  basicstyle=\ttfamily\footnotesize,
  breaklines=true,
  breakatwhitespace=false,
  columns=fullflexible,
  keepspaces=true,
  showstringspaces=false,
  frame=single,
  xleftmargin=0.5em,
  framexleftmargin=0.5em,
  aboveskip=0.8em,
  belowskip=0.4em
}

\lstdefinestyle{cottext}{
    basicstyle=\small\ttfamily,
    numbers=none,
    frame=single,
    backgroundcolor=\color{yellow!5},
    xleftmargin=2em,
    xrightmargin=2em,
    breaklines=true,
    breakatwhitespace=true,
    aboveskip=1em,
    belowskip=1em,
    columns=flexible,
    keepspaces=true
}

\usepackage{lineno}
\usepackage{lipsum}
\usepackage{pifont} 
\usepackage[table]{xcolor}

\definecolor{darkblue}{rgb}{0, 0, 0.5}
\hypersetup{colorlinks=true, citecolor=darkblue, linkcolor=darkblue, urlcolor=darkblue}

\title{Learning Reasoning World Models for Parallel Code}

\author{
Gautam Singh$^{1}$, \textbf{Arjun Guha}$^{2}$, \textbf{Bhavya Kailkhura}$^{1}$ \textbf{\&} \textbf{Harshitha Menon}$^{1}$\\\\
\hspace{0.25em}$^{1}$\textit{Lawrence Livermore National Laboratory}\\
\hspace{0.7em}\texttt{\{singh68, gopalakrishn1\}@llnl.gov}\\\\
\hspace{0.25em}$^{2}$\textit{Northeastern University}\\
\hspace{0.7em}\texttt{a.guha@northeastern.edu}
}

\begin{document}

\ifcolmsubmission
\linenumbers
\fi

\maketitle

\begin{abstract}
Large language models have shown remarkable ability in serial code generation, but they still struggle with parallel code for which training data is comparatively scarce. A common remedy is to use coding agents that interact with external tools, but tool calls can be costly and sometimes impractical, e.g., for partially written code. We propose \textit{Parallel-Code World Models (PCWMs)}, reasoning LLMs that aim to predict tool outcomes directly from parallel source code. To train PCWMs, we design a novel exploration and data generation pipeline that samples diverse parallel-coding problems and candidate implementations across multiple domains, then executes them via tools to record data races and performance profiles. From these, we synthesize reasoning traces that causally connect source code to observed tool outcomes. 
Fine-tuning on the resulting data yields noticeable gains, with a 7B-parameter world model improving from 64.3\% to 72.8\% accuracy in race-outcome prediction, while an 8B-parameter model improves in a performance profiling task from 49.3\% to 58.6\% accuracy.
Furthermore, when open-weight models were tasked with fixing data races, world-model feedback improved their race-fixing rates relative to self-feedback by 2.7\%-9.1\% using our 7B-parameter world model and by 6.1\%-11.1\% using our 14B-parameter world model. Our results suggest that reasoning world models may have the potential to serve alongside external tool calls in parallel-coding agents.
\end{abstract}

\section{Introduction}

Large language models (LLMs) have transformed code development through code generation, explanation, and debugging workflows \citep{codex, codellama, chatgpt, yang2024sweagent}. However, much of this progress has been concentrated in data-rich languages, programming models, and frameworks. Many real-world applications depend on niche and specialized frameworks with scarce training data, where language models struggle to produce correct, well-optimized code~\citep{chen2024landscape, mora2024synthetic}.

A setting that is both important and representative of this scarcity problem is \textit{parallel programming}. In parallel programming, a code's functionality is decomposed across multiple threads or processes that execute concurrently---giving rise to complex non-deterministic scheduling, synchronization, and communication dynamics. Because high-quality parallel-code data is comparatively scarce, LLMs struggle more on parallel programming tasks than on serial programming tasks \citep{pareval}. A natural way to address this is via coding agents that can plan and search programs while obtaining feedback from the environment via relevant tools. However, real tool calls can be costly to configure and can sometimes be impractical, e.g., for partially written code snippets.

To mitigate external tool calls for parallel codes, there are two relevant lines of research. First, in model-based reinforcement learning, agents use learned world models to plan in lieu of expensive environment interactions \citep{ha2018worldmodels,hafner2020dreamer}. Recent work adapts this idea via code world models (CWMs), where LLMs are trained to model execution behavior and traces \citep{copet2025cwm,armengol2025execute}. However, the notion of code world models remains underexplored for \emph{parallel code}. Second, in parallel-computing research, neural surrogate models have been used since the early days of deep learning to predict properties of interest from parallel programs \citep{malakar2018benchmarking,marathe2017deeptransfer,chen2018learningtensor,cummins2021programl,nichols2024crossarch, bolet2025llmperf, chunhua}. Yet this line of work is yet to fully explore the reasoning ability of large language models \textit{to model the process of arriving at tool outcomes from source code alone}.

In this work, we propose \textit{Parallel-Code World Models (PCWMs)}, reasoning large language models that emulate outcomes from analysis tools for parallel code. We show that this task is difficult without explicit reasoning supervision because of the complex causal dynamics in parallel execution. Consequently, we introduce a novel exploration and data-collection pipeline that gathers diverse parallel codes, tool outcomes, and reasoning traces, yielding a reasoning dataset for training parallel-code world models.

The contributions of this paper are: (1) we introduce {Parallel-Code World Models (PCWMs)}, reasoning LLMs that emulate outcomes of tools relevant to parallel coding directly from source code;
(2) we design a scalable exploration and data-collection pipeline spanning diverse domains, problems, and parallelization strategies, and we build datasets for two parallel-coding focused tools: \textit{ThreadSanitizer} for race detection and \textit{Caliper} for work-percentage profiling;
(3) we show that emulating these tools is challenging without explicit reasoning supervision, and therefore collect chains of thought that causally connect source code to observed tool outcomes; and
(4) we show that trained PCWMs can provide useful feedback for downstream bug-fixing, e.g., fixing data races, and produce noticeable gains on benchmark evaluations.

\begin{figure}
    \centering
    \includegraphics[width=0.98\linewidth]{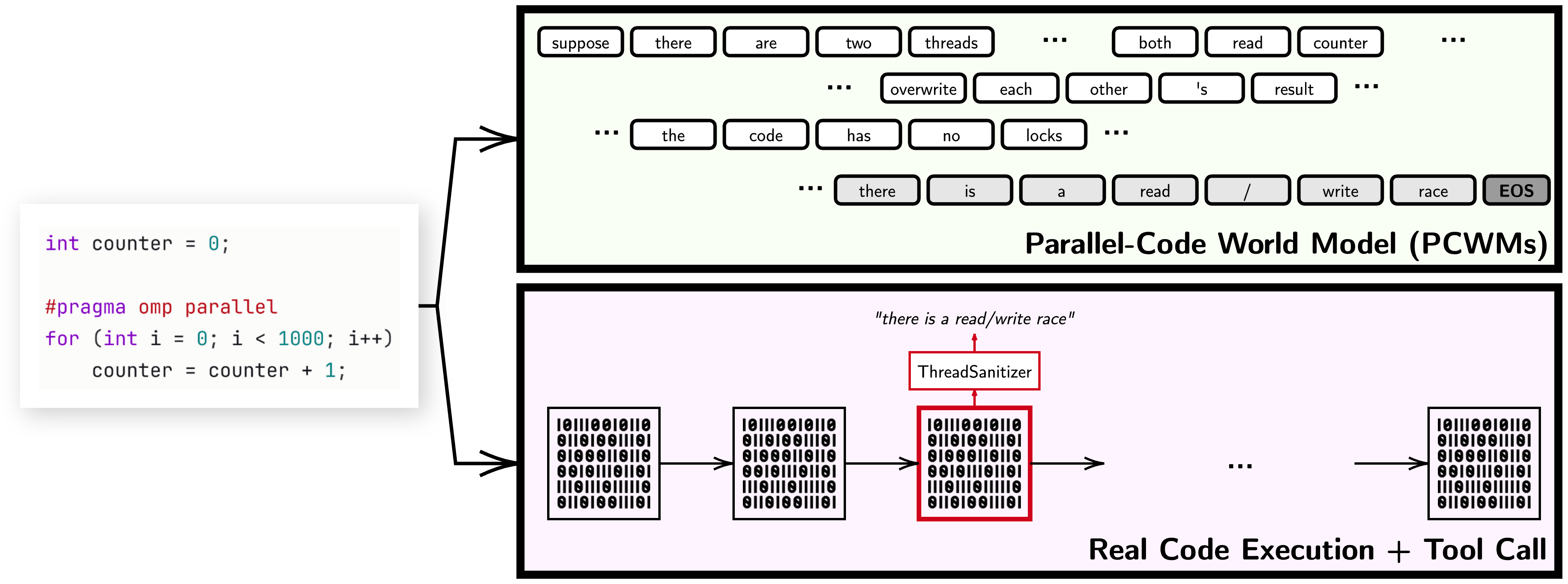}
    \caption{\textbf{Overview.} We illustrate the contrast between two ways an agent may receive feedback about code. \textbf{Top:} \textit{Parallel-Code World Models (PCWMs)} aim to simulate the execution process causally in terms of higher-level concepts and events to infer the would-be outcome of an external tool. It does so via an auto-regressive chain of reasoning tokens. The reasoning tokens are illustrated with white tokens, and the outcome tokens are illustrated with gray tokens. \textbf{Bottom:} Real tool requires precise execution of the code, which can be costly to configure and may be impractical for partial code snippets.}
    \label{fig:one}
\end{figure}

\section{PCWM: Parallel-Code World Model}
Parallel-Code World Models (PCWMs) are reasoning models that take a parallel program source code $\bx_\texttt{code}$ and predict a reasoning trace $\bz$ and an expected execution outcome $\by$.
Within the reasoning trace, it performs high-level thought experiments and analysis of thread interactions, synchronization behavior, and possible failure modes.
Formally, we model this as sampling $(\bz, \by)$ conditioned on code $\bx_\texttt{code}$ as follows:
\begin{align*}
    \bz, \by \leftarrow \texttt{PCWM}(\bx_\texttt{code}),
\end{align*}
where $\texttt{PCWM}$ denotes a parallel-code world model. In practice, we want the world model to be a reasoning LLM that can produce $\bz, \by$ effectively, quickly, and cheaply.
For an agent calling the world model, this would produce a favorable cost-benefit trade-off for rolling out a world model over setting up and executing true code and invoking the true tool.

\subsection{Data Collection}
In this section, we describe the pipeline used to construct our training dataset of parallel-code, tool outcomes, and reasoning traces.
\begin{figure*}[t]
    \centering

    \begin{subfigure}[t]{0.98\textwidth}
        \centering
        \caption{Generate Problem}
        \begin{lstlisting}[style=tooloutput,basicstyle=\small\ttfamily]
Given a large vector of size n, compute the dot product of this vector with another vector of the same size. Input: vectors v and w. Output: scalar value\end{lstlisting}
    \end{subfigure}


    \begin{minipage}[t]{0.49\textwidth}
        \centering
        \begin{subfigure}[t]{\textwidth}
            \centering
            \caption{Generate Harness}
            \begin{lstlisting}[style=cppcode,basicstyle=\tiny\ttfamily]
bool validate() {
    const size_t numTries = 10;
    const size_t vectorSize = 1000;
    std::random_device rd;
    std::mt19937 gen(rd());
    std::uniform_real_distribution<> dis(
        -100.0, 100.0);

    for (int i = 0; i < numTries; i++) {
        // Generate random test vectors
        std::vector<double> v(vectorSize);
        std::vector<double> w(vectorSize);
        for (size_t j = 0; j < vectorSize; ++j) {
            v[j] = dis(gen);
            w[j] = dis(gen);
        }

        // Compare reference vs generated
        double correctResult = reference(v, w);
        double testResult = generated(v, w);

        bool isCorrect =
            std::abs(correctResult - testResult)
            < 1e-6;

        if (!isCorrect) return false;
    }
    return true;
}
            \end{lstlisting}
        \end{subfigure}
    \end{minipage}
    \hfill
    \begin{minipage}[t]{0.48\textwidth}
        \centering
        \begin{subfigure}[t]{\textwidth}
            \centering
            \caption{Generate Reference Implementation}
            \begin{lstlisting}[style=cppcode,basicstyle=\scriptsize\ttfamily]
double reference(
    const std::vector<double>& v,
    const std::vector<double>& w) {
    double dotProduct = 0.0;
    for (size_t i = 0; i < v.size(); ++i) {
        dotProduct += v[i] * w[i];
    }
    return dotProduct;
}
            \end{lstlisting}
        \end{subfigure}

        \vspace{0.3cm}

        \begin{subfigure}[t]{\textwidth}
            \centering
            \caption{Generate Candidate Implementation}
            \begin{lstlisting}[style=cppcode,basicstyle=\scriptsize\ttfamily]
double generated(
    const std::vector<double>& v,
    const std::vector<double>& w) {
    double dotProduct = 0.0;
    #pragma omp parallel for
    for (size_t i = 0; i < v.size(); ++i) {
        dotProduct += v[i] * w[i];
    }
    return dotProduct;
}
            \end{lstlisting}
        \end{subfigure}
    \end{minipage}


    \begin{subfigure}[t]{0.98\textwidth}
    \centering
    \caption{Generate CoT Conditioned on Tool Outcome}
    \begin{lstlisting}[style=cottext,basicstyle=\small\ttfamily]
... When each thread does the addition to dotProduct, they might be reading and writing the same memory location without synchronization. That's a classic case for a data race. The critical part is the 'dotProduct += ...' operation: first, the current value of dotProduct is read, then the product is added, and the new value is written back ...\end{lstlisting}
\end{subfigure}
    \caption{\textbf{Illustration of Our LLM-Driven Reasoning Data Generation Pipeline.}
    Shown is one instance produced from our LLM-driven data-generation pipeline: (a) a parallel programming problem statement, (b) a generated harness, (c) a sequential reference implementation, (d) a candidate OpenMP parallel implementation, and (e) an example chain-of-thought generated after observing the tool outcome for the candidate code.}
    \label{fig:problem-reference-example}
\end{figure*}

\textbf{Problem and Harness Generation.} We first prompt an LLM to generate a parallel programming problem and a corresponding harness code. Here: (1) a parallel programming problem refers to a programming task whose code solution requires parallelism for efficient execution, e.g., matrix multiplication with multiple workers; (2) the harness code specifies the execution scaffold for the task that includes what inputs are provided to the candidate code and how outputs are returned or checked. It therefore defines the interface and runtime context in which candidate code is executed.
Conditioned on the problem and the harness, we then sample a diverse set of candidate parallel-code implementations.
\begin{align*}
    \bx_\texttt{problem}, \bx_\texttt{harness} \leftarrow \texttt{LLM}(\bc_\texttt{task}) && \Longrightarrow && \bx_{\texttt{code}} \leftarrow  \texttt{LLM}(\bx_\texttt{problem}, \bx_\texttt{harness}, \bc_\texttt{implement}).
\end{align*}
Here, $\bx_\texttt{problem}$ denotes the parallel programming problem statement, $\bx_\texttt{harness}$ the corresponding harness code, and $\bx_\texttt{code}$ a candidate parallel-code implementation. Moreover, $\texttt{LLM}(\cdot)$ denotes LLM-based generation, while $\bc_{\texttt{task}}$ and $\bc_{\texttt{implement}}$ denote the context or prompts used for problem/harness generation and candidate-code generation, respectively.

\textbf{Exploration Challenge.} A key challenge in the above pipeline is ensuring sufficient exploration during both problem generation and candidate-code generation. If sampled problems are overly concentrated in a narrow domain, the resulting world model will also be narrow in its capability. Similarly, it is important to obtain diverse candidate code implementations for each problem to expose distinct parallelization strategies. In this work, we address this by enlisting 10 domains relevant for parallel programming, defining 8 seed problems per domain, and prompting an LLM to create 20 variants of each seed problem. For generating code candidates, we prompt the LLM to produce 4 implementations per problem that are distinct from one another in their parallelization strategies. In our setup, we use a strong model, i.e., \texttt{GPT-4o}, for problem and harness generation. For candidate parallel-code generation, we use a zoo of open-weight models to improve code diversity: \texttt{Llama-3.3-70B}, \texttt{Llama-3.1-8B}, \texttt{Gemma-3-27B}, and \texttt{Phi-4}. Furthermore, we take OpenMP to be our parallelism framework.

\textbf{Tool Outcome and CoT Generation.} We then run each candidate code $\bx_\texttt{code}$ using its harness while applying the desired toolchain to obtain the tool outcome $\by$. In this work, we consider two tools: \textit{ThreadSanitizer} and \textit{Caliper} that we discuss in more detail in Section~\ref{sec:tools}. Given the candidate code $\bx_\texttt{code}$ and its observed tool outcome $\by$, we sample a chain of thought $\bz$ to serve as a reasoning trace for world-model learning. This process can be described as follows:
\begin{align*}
    \by \leftarrow \texttt{Tool}(\bx_\texttt{code}, \bx_\texttt{harness}) && \Longrightarrow && \bz \leftarrow \texttt{LLM}(\bx_\texttt{code}, \by, \bc_\texttt{cot}).
\end{align*}
Here, $\bc_{\texttt{cot}}$ denotes the prompt used for CoT synthesis. All prompts are provided in Appendix~\ref{app:prompts}. In designing the prompt $\bc_{\texttt{cot}}$, we instruct the teacher that the provided outcome $\by$ is not a foregone fact to analyze, but a target to be reached by reasoning forward causally from the code $\bx_\texttt{code}$ to the observed outcome. 
In our experiments, we compare two teacher models: \texttt{QwQ-32B} and \texttt{GPT-OSS-20B}; and evaluate their usefulness as generators of CoTs for world-model learning.

\textbf{Dataset.} By repeating the above pipeline across many sampled problems, multiple candidate codes per problem, tool executions, and multiple CoT samples, we collect a large set of training tuples $\cD = \{(\bx_\texttt{code}, \bz, \by)\}$. Concretely, each tuple contains a candidate code $\bx_\texttt{code}$, its tool outcome $\by$, and a corresponding CoT $\bz$. In this work, we collect 27K data samples for each tool we consider for conducting reasoning post-training.

\subsection{Tools}
\label{sec:tools}
We consider the following parallel-coding relevant tools for our data-generation pipeline to collect data for world model training. Outcomes from these tools are illustrated in Figure \ref{fig:tool-outcomes}.

\textbf{ThreadSanitizer~\citep{threadsanitizer}.} It is an analysis tool that detects data races and other threading errors in multithreaded programs at runtime.
We run ThreadSanitizer on each candidate program execution to detect data races at runtime. Specifically, for each candidate code $\bx_\texttt{code}$, it produces the outcome $\by$, which is a list of races, where each race is described by the affected line or variable and the race type. The world model trained on the data generated using ThreadSanitizer outcomes is called \textit{ThreadSanitizer World Model}.

\begin{figure*}[t]
    \centering

    \begin{subfigure}[t]{0.42\textwidth}
        \centering
        \caption{ThreadSanitizer Outcome}
        \begin{lstlisting}[style=tooloutput]
        
[
  {
    'type': 'read/write race',
    'code_locations': [
      'generated.cc:14'
    ]
  },
  {
    'type': 'write/write race',
    'code_locations': [
      'generated.cc:14'
    ]
  }
]\end{lstlisting}
    \end{subfigure}
    \hfill
    \begin{subfigure}[t]{0.56\textwidth}
        \centering
        \caption{Caliper Outcome}
        \begin{lstlisting}[style=tooloutput]
For code snippet:
#pragma omp for
for (size_t i = 0; i < n; ++i) {
    for (size_t j = 0; j < n; ++j) {
        for (size_t k = 0; k < n; ++k) {
            C[i][j] += temp[i][k] * A[k][j];
        }
    }
}

Caliper measures:
- For 4 threads, a work percentage of 96
- For 16 threads, a work percentage of 76
- For 64 threads, a work percentage of 44
- For 128 threads, a work percentage of 47\end{lstlisting}
    \end{subfigure}

    \caption{\textbf{Illustration of Tool Outcomes.} The left panel shows a representative ThreadSanitizer report identifying data races in the candidate OpenMP program. The right panel shows a representative Caliper profile reporting the work percentages of a parallel code region across different thread counts, reflecting how efficiently execution time is spent on useful work rather than OpenMP overhead.}
    \label{fig:tool-outcomes}
\end{figure*}

\textbf{Caliper~\citep{caliper}.} It is a performance-profiling tool for HPC programs that records fine-grained runtime metrics (e.g., compute work vs. synchronization overhead) across code regions.
We use Caliper to obtain the work percentages of parallel regions in each program.
Here, work percentage is the fraction of total OpenMP thread time spent doing useful application work, rather than OpenMP runtime overhead, e.g., barriers, synchronization, and idle time. The world model trained on the data generated using Caliper work-percentage outcomes is called \textit{Caliper World Model}. For Caliper, measurements for the same/similar source code can show variability due to outside factors, e.g., characteristics of the execution environment. Hence, for Caliper, we seek to jointly model Caliper measurements (obtained under similar conditions) for a \textit{pair} of source codes, in order to model their relative structure:
\begin{align*}
    \by^{(i)} \leftarrow \texttt{Tool}(\bx_\texttt{code}^{(i)}, \bx_\texttt{harness}) && \Longrightarrow && \bz^{(ij)} \leftarrow \texttt{LLM}(\bx_\texttt{code}^{(i)}, \bx_\texttt{code}^{(j)}, \by^{(i)}, \by^{(j)}, \bc_\texttt{cot}).
\end{align*}
where $\bz^{(ij)}$ is a CoT that analyzes the relative differences between the two codes $\smash{\bx_\texttt{code}^{(i)}, \bx_\texttt{code}^{(j)}}$ in order to reach a joint prediction of their Caliper measurements $\smash{\by^{(i)}, \by^{(j)}}$.

\subsection{World Model Training}
\label{sec:wm-training}
We concatenate the CoT tokens and the outcome tokens into a token-sequence $[\bz, \by]$, and we perform supervised fine-tuning of the world model LLM to learn to predict $[\bz, \by]$ given a parallel code $\bx_\texttt{code}$. We then minimize the following loss function via supervised fine-tuning: $\cL(\ta_\texttt{PCWM}) = -\log p_{\ta_\texttt{PCWM}}([\bz, \by] \mid \bx_\texttt{code})$. This is equivalent to supervised fine-tuning (SFT) with a completion-only loss over the expected assistant response $[\bz, \by]$. For Caliper, the world model jointly takes as input a pair of codes $\smash{\bx^{(i)}, \bx^{(j)}}$ and the fine-tuning targets are a concatenation of the CoT tokens and Caliper outcomes for both codes: $\smash{[\bz^{(ij)}, \by^{(i)}, \by^{(j)}]}$. The training loss is:
$\smash{\cL(\theta_{\texttt{PCWM}}) = -\log p_{\theta_{\texttt{PCWM}}}\!\left([\bz^{(ij)}, \by^{(i)}, \by^{(j)}] \mid \bx_{\texttt{code}}^{(i)}, \bx_{\texttt{code}}^{(j)}\right).}$

\subsection{Applications of PCWMs}
\label{sec:applications}

\textbf{Parallel Code Understanding.} A world model can help human developers quickly spot likely performance bugs in partially written parallel-code snippets, even before full integration or execution.
It can also support code review by surfacing likely synchronization bottlenecks, race-prone regions, and other parallelization-related bugs.

\textbf{Feedback Providing Tool for Parallel-Code Agent.} In a parallel-coding agent, the world model can be used as a source of feedback and a surrogate for expensive or slow tool calls during iterative refinement. We describe a world-model-guided bug-fixing loop for a parallel-code agent as follows, which we will use in experiments to evaluate the PCWMs:
\begin{align*}
[\bz, \by] &\leftarrow \texttt{PCWM}(\bx_{\texttt{code}}); \quad  \bee \leftarrow \texttt{LLM}(\bx_{\texttt{code}}, \bz, \by, \bc_{\texttt{edit}}); \quad \bx'_{\texttt{code}} \leftarrow \texttt{LLM}(\bx_{\texttt{code}}, \bee, \bc_{\texttt{apply}});
\end{align*}
where $\smash{\bx_{\texttt{code}}}$ denotes a potentially buggy code, and $\smash{\bz}$ and $\smash{\by}$ denote the predicted reasoning trace and tool outcome, respectively. Moreover, $\smash{\bee}$ denotes the edit proposal generated from world-model feedback, while $\smash{\bc_{\texttt{edit}}}$ and $\smash{\bc_{\texttt{apply}}}$ denote the prompts or contexts used for edit generation and code updating, respectively. The output of the refinement step is the updated candidate code $\smash{\bx'_{\texttt{code}}}$.

\section{Related Work}
\textbf{Code World Models and Feedback-Driven Code Agents.} In deep learning, RL agents can improve their decisions by planning against a neural model of environment dynamics \citep{ha2018worldmodels,hafner2020dreamer}. Recent work brings this idea to code agents: training world models on code execution-trace data \citep{copet2025cwm, armengol2025execute}. While \cite{copet2025cwm} discuss the idea of translating structured execution traces into a free-form natural language chain of thoughts, to our knowledge, the focus of their work is not on parallel code and related tools/profilers, unlike ours. Yet the notion of world models is yet to be explored for HPC settings, e.g., for parallel code generation. Simultaneously, today's code-generation systems rely on feedback loops \citep{le2022coderl, chen2023selfdebug, shinn2023reflexion, gehring2024rlef}. Frameworks such as SWE-agent and OpenHands utilize actor--environment interaction over repositories and tools \citep{yang2024sweagent,wang2024openhands}. In HPC-specific contexts, HPC-Coder adapts LLMs for parallel-code tasks, including performance-related predictions \citep{nichols2024hpccoder}. Although these show the utility of feedback for coding agents, they rely on external executions. In contrast, our work explicitly trains reasoning LLMs to imitate analysis tools by mapping code directly to a causal reasoning trace and an expected tool outcome.

\textbf{Neural Surrogates in HPC.} There is deep interest in the community to utilize large language models for HPC tasks \citep{pareval, nichols2024performance, godoy2023evaluation, godoy2024large, valero2023comparing, joel2024survey, teranishi2025leveraging, nader2025llm}. Since the early days of deep learning, neural surrogate models for performance/cost to enable offline search and optimization have been explored \citep{liu2021gptune, malakar2018benchmarking, marathe2017deeptransfer, nichols2024crossarch, chen2018learningtensor,cummins2021programl}. Works on LLM-based performance inference include \cite{bolet2025llmperf, chunhua}. However, these approaches mainly focus on scalar labels and do not leverage reasoning as a means to predict detailed tool outcomes from source code alone.

\section{Experiments}

\subsection{ThreadSanitizer World Model for Data Race Analysis}

\subsubsection{Setup}
To evaluate the ThreadSanitizer world model trained in Section~\ref{sec:wm-training}, we use DataRaceBench (DRB), a benchmark suite for data-race detection in shared-memory parallel programs, with a focus on OpenMP. DRB consists of approximately 200 OpenMP codes containing race-free and intentionally racy examples. We use DRB in two settings: (1) intrinsic evaluation, where the world model predicts whether a given DRB program contains a race condition, and (2) agentic race fixing, where a race-fixing agent edits DRB programs and uses the ThreadSanitizer world model as a feedback source, as described in Section~\ref{sec:applications}.

\subsubsection{Results}

\textbf{Importance of Teaching to Think.} Table~\ref{tab:teaching-to-think} shows that, on average, base models benefit from being prompted to think before answering with a list of potential data races. Supervised fine-tuning on outcome-only data, $\cD_\texttt{SFT-No-CoT} = \{(\bx_\texttt{code}, \by)\}$, further improves performance over the base models. Additional gains are obtained by fine-tuning on CoT-augmented data, $\cD_\texttt{SFT-With-CoT} = \{(\bx_\texttt{code}, \bz, \by)\}$. These results show that: (1) Although prompting to think is helpful, it alone is not enough. (2) We need to explicitly \textit{train} the models on reasoning behaviors in order for them to serve as better world models for parallel-codes.

\textbf{Effect of CoT Teacher Model.} We analyze the effect of the choice of the teacher model used to synthesize the CoTs. We compare 2 teacher models: \texttt{GPT-OSS-20B} and \texttt{QwQ-32B}. In Table~\ref{tab:teaching-to-think}, we can see that \texttt{QwQ-32B} performs better than \texttt{GPT-OSS-20B} for synthesizing chains of thought---achieving on average 73.0\% vs. 69.6\%, respectively. That being said, there appears to be a teacher--student compatibility effect: \texttt{Llama-3.1-8B} appears to leverage \texttt{GPT-OSS-20B} chains better than \texttt{Qwen-2.5-7B}, achieving 69.6\% versus 66.5\%, respectively.

\begin{table}[t]
\centering
\caption{\textbf{Importance of Teaching to Think for ThreadSanitizer World Model.} We report the accuracy of predicting the presence of a race condition on OpenMP codes in DataRaceBench. We compare the world models with thinking modes \textit{on} and \textit{off}. For the \textit{+ SFT} setting, we also indicate the CoT teacher model used. The accuracies are computed using 16 samples per prompt with temperature 0.6.}\label{tab:teaching-to-think}
\begin{tabular}{@{}l|cc|c|cc@{}}
\toprule
& \multicolumn{2}{c}{Base} & \multicolumn{3}{|c}{+ SFT} \\
\cmidrule(lr){2-3} \cmidrule(lr){4-6}
& & & \multicolumn{1}{c}{ } & \multicolumn{2}{|c}{CoT Teacher} \\
\cmidrule(lr){5-6}
World Model & No CoT & With CoT & No CoT & \texttt{GPT-OSS-20B} & \texttt{QwQ-32B} \\
\midrule
\texttt{Qwen-2.5-32B} & 67.0\% & 66.4\% & 70.5\% & 72.4\% & \textbf{75.2\%} \\
\texttt{Qwen-2.5-14B} & 62.2\% & 67.1\% & 67.8\% & 69.7\% & \textbf{74.2\%} \\
\texttt{Qwen-2.5-7B}  & 58.5\% & 64.3\% & 64.4\% & 66.5\% & \textbf{72.8\%} \\
\texttt{Llama-3.1-8B} & 52.6\% & 53.9\% & 64.2\% & 69.6\% & \textbf{69.7\%} \\ \midrule
\rowcolor{gray!10} Average & 60.1\% & 62.9\% & 66.7\% & 69.6\% & \textbf{73.0\%} \\
\bottomrule
\end{tabular}
\end{table}

\textbf{Importance of Training Dataset Size.} We analyze the effect of the dataset size that was used to train the ThreadSanitizer-mimicking world model. In Table~\ref{tab:cot-hindsight-and-scaling}, on average, increasing the SFT dataset size helps improve performance, suggesting that there is value in designing better exploration policies for generating problems and candidate codes to construct larger training datasets. 
Additionally, larger student models appear less sensitive to CoT dataset size: for \texttt{Qwen-2.5-32B} trained on \texttt{QwQ-32B} CoTs, performance gains are relatively stable across 6K, 13K, and 27K samples (76.0\%, 74.9\%, 75.2\%), whereas \texttt{Qwen-2.5-7B} shows a monotonic improvement from 6K to 27K samples (66.2\% to 72.8\%).

\textbf{Utilizing the World Model as a Feedback Source for Race Fixing.} In Table~\ref{tab:race-fixing-open} and \ref{tab:race-fixing-commercial}, we report the percentage of race-free codes after one pass of the race-fixing loop on DataRaceBench. We observe the following: (1) world models after SFT provide more effective feedback than base world models, leading to higher race-fixing performance in general; (2) larger world models generally provide better performance boost than smaller ones; and (3) on average, world models after SFT are more effective feedback providers than open-weight actor models leveraging self-feedback during race-fixing; (4) Interestingly, for open-weight actor models, world model feedback from world models 14B or larger in size even outperforms oracle or ground-truth race feedback. We think this is because world models don't just provide the final tool outcomes but also the causal process via which those outcomes are reached, thus further helping the race-fixing task. (5) Finally, when testing commercial models as race-fixing actor models, we find that these tend to be less affected by feedback in general and do not gain significantly from feedback, including oracle feedback. 

\textbf{Compute Cost of World Model Calls.} In Table~\ref{tab:pcwm-compute}, we analyze the compute cost of querying world models for feedback. We obtain these measurements by evaluating each model on DataRaceBench programs and reporting per-response FLOPs, average response length, and accuracy. We can see that: (1) the average response length needed to reach the data-race prediction tends to drop with increasing model size, with lengths decreasing from 5.2K tokens to 3K tokens when going from 7B to 32B parameter models. (2) Because of this effect, although larger models may appear more expensive, their stronger reasoning ability can allow them to reach the correct answer with fewer generated tokens. Hence, larger models may achieve higher accuracy while needing similar FLOPs/response as smaller models.

\begin{table*}[t]
\centering
\caption{\textbf{Reasoning Training Dataset Analysis.} 
We analyze the effect of the number of reasoning data samples used for world model training. All reported numbers are accuracies of predicting the presence of a race condition on OpenMP codes in DataRaceBench.}
\label{tab:cot-hindsight-and-scaling}
\vspace{0.3em}
\begin{subtable}[t]{0.67\textwidth}
\centering
\setlength{\tabcolsep}{3pt}
\begin{tabular}{@{}l|c|c|ccc@{}}
\toprule
     &  &  & \multicolumn{3}{c}{SFT Dataset Size} \\ \cmidrule(l){4-6}
World Model & Base & CoT Teacher & 6K & 13K & 27K \\ \midrule
\multirow{2}{*}{\texttt{Qwen-2.5-32B}} & \multirow{2}{*}{66.4\%} & \texttt{GPT-OSS-20B} & 69.8\% & \underline{70.8\%} & \textbf{72.4\%} \\
 &  & \texttt{QwQ-32B} & \textbf{76.0\%} & 74.9\% & \underline{75.2\%} \\ \midrule
\multirow{2}{*}{\texttt{Qwen-2.5-7B}} & \multirow{2}{*}{64.3\%} & \texttt{GPT-OSS-20B} & 58.9\% & \underline{63.7\%} & \textbf{66.5\%} \\
 &  & \texttt{QwQ-32B} & 66.2\% & \underline{69.7\%} & \textbf{72.8\%} \\ \midrule
\rowcolor{gray!10}
\multicolumn{3}{r|}{Average} & 67.7\% & 69.8\% & \textbf{71.7\%} \\
\bottomrule
\end{tabular}
\end{subtable}
\end{table*}

\begin{table}[t]
\centering
\caption{\textbf{Race-Fixing using Open-Weight Actor Models on DataRaceBench.} We task an LLM to act as a race-fixing agent to fix race conditions in OpenMP codes in DataRaceBench. The race-fixing agent receives race analysis and detection feedback from the world model. The race-free percentages of the fixed codes are reported as measured using ThreadSanitizer.}
\label{tab:race-fixing-open}
\setlength{\tabcolsep}{7pt}
\begin{tabular}{@{}l|c|c|cc|cc|cc@{}}
\toprule
 & \multicolumn{8}{c}{Feedback Source} \\ \cmidrule{2-9}
 & \multirow{2}{*}{Oracle} & \multirow{2}{*}{Self} & \multicolumn{6}{c}{World Model} \\
\cmidrule(l){4-9}
Actor Model &  &  & \multicolumn{2}{c|}{\texttt{Qwen-2.5-32B}} & \multicolumn{2}{c|}{\texttt{Qwen-2.5-14B}} & \multicolumn{2}{c}{\texttt{Qwen-2.5-7B}} \\
\cmidrule(lr){4-5}\cmidrule(lr){6-7}\cmidrule(lr){8-9}
 &  &  & Base & +SFT & Base & +SFT & Base & +SFT \\
\midrule
\texttt{Llama-3.1-70B} & 79.4\% & 73.8\% & 75.3\% & \textbf{80.8\%} & 76.3\% & \textbf{78.3\%} & 72.4\% & \textbf{75.8\%} \\
\texttt{Gemma-3-27B}   & 69.3\% & 63.9\% & 66.1\% & \textbf{73.0\%} & 66.3\% & \textbf{71.0\%} & 66.0\% & \textbf{69.7\%} \\
\midrule
\rowcolor{gray!10} Average & 74.4\% & 68.9\% & 70.7\% & \textbf{76.9\%} & 71.3\% & \textbf{74.7\%} & 69.2\% & \textbf{72.8\%} \\
\bottomrule
\end{tabular}
\end{table}
\begin{table}[t]
\centering
\setlength{\tabcolsep}{3pt}
\caption{\textbf{Compute Cost Analysis of World Models after SFT.} We compare the computational cost and predictive performance of different world models after SFT on DataRaceBench. For each model, we report the estimated FLOPs per response, average generated sequence length, and accuracy for predicting the presence of a race condition in OpenMP code.}
\label{tab:pcwm-compute}
\setlength{\tabcolsep}{5.5pt}
\begin{tabular}{@{}l|cccc@{}}
\toprule
Metric & \texttt{Qwen-2.5-32B} & \texttt{Qwen-2.5-14B} & \texttt{Qwen-2.5-7B} & \texttt{Llama-3.1-8B} \\
\midrule
Model Size & 32B & 14B & \textbf{7B} & 8B \\
TFLOPs/Response & \textbf{2183.76} & 3093.78 & 2346.80 & 3066.34 \\
Average Response Length & \textbf{3049.84} & 3915.18 & 5216.02 & 5282.77 \\
\midrule
Post-SFT Accuracy & \textbf{75.2\%} & 74.2\% & 72.8\% & 69.7\% \\
\bottomrule
\end{tabular}
\end{table}

\subsection{Caliper World Model for Relative Work Percentage Analysis}

\textbf{Setup.} To evaluate the Caliper world model trained in Section~\ref{sec:wm-training}, we take ParEval \citep{pareval}, a benchmark consisting of 54 OpenMP programming problems. For each problem, we generate multiple distinct OpenMP solutions for each problem using \texttt{GPT-4.1-mini} and \texttt{GPT-5.1}, and filter for the correct solutions. We then profile each correct implementation using Caliper at thread counts 4, 16, 64, and 128. This collected data serves as a testbed for evaluating our Caliper world model. For each problem in ParEval, we take pairs of distinct OpenMP solutions for the same problem, and we roll out the Caliper world model to predict Caliper work percentages for the code-pair. We then rank the codes in the code-pair based on the world model-predicted work-percentages and test whether the ranking matches that based on true Caliper measurements. In Table \ref{tab:caliper-wm}, we show the accuracy of their agreement.

\begin{figure}[t]
    \centering
    \includegraphics[width=0.99\linewidth]{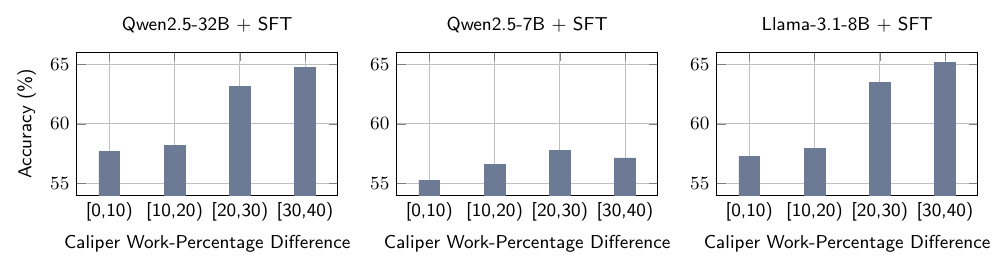}
    \caption{\textbf{Effect of Caliper Measurement Difference on World Model Accuracy.} We report the accuracy (averaged over thread counts) of predicting rank ordering of OpenMP code pairs in terms of which code will exhibit higher work percentage as per Caliper. We disentangle the effect of possible stochasticity in rank ordering when the work-percentage difference as measured by Caliper is small (e.g., a difference of less than 10\% versus a difference of 30-40\%) by reporting accuracy for various work-percentage difference ranges. This is done on solutions for ParEval problems obtained using \texttt{GPT-4.1-mini}. Both Caliper world models show better capability of evaluating the code-pair when the work percentage difference in the evaluated code-pair is larger.}
    \label{fig:ambiguity}
\end{figure}

\textbf{Results.} Table~\ref{tab:caliper-wm} shows the following: (1) Supervised fine-tuning consistently improves accuracy for the world models across nearly all thread counts. On average across models, accuracy increases from 46.4\% to 58.0\% when profiling \texttt{GPT-4.1-mini}'s ParEval implementations, and from 46.0\% to 53.7\% when profiling \texttt{GPT-5.1}'s ParEval implementations. The gains from fine-tuning for \texttt{Qwen-2.5-7B} are smaller compared to those for \texttt{Llama-3.1-8B} and \texttt{Qwen-2.5-32B}. (2) World modeling performance on OpenMP code generated by \texttt{GPT-4.1-mini} is better than that on code generated by \texttt{GPT-5.1}. A likely explanation is that \texttt{GPT-4.1-mini}'s OpenMP code is simpler and more similar to the distribution of code seen during world model training. This highlights the need to improve the exploration aspect of the data synthesis pipeline in future work. Another noteworthy observation, illustrated in Figure~\ref{fig:ambiguity}, is that the agreement between the Caliper world model and ground-truth Caliper measurements improves as the measured work-percentage gap between two codes increases. This suggests that world models struggle more with near-tie cases than with cases where measurement differences are larger and potentially less ambiguous.

\begin{table}[t]
    \centering
    \small
    \caption{\textbf{Caliper World Model for Ranking Parallel-Codes in terms of Work-Percentage.} We report the accuracy of predicting the rank ordering for OpenMP code-pairs in terms of which code will exhibit higher work percentage as per Caliper. The OpenMP code-pairs are obtained by getting verified OpenMP implementations for problems in the ParEval benchmark using \texttt{GPT-5.1} and \texttt{GPT-4.1-mini}. We compare world modeling performance of base models against the fine-tuned models for thread counts 4, 16, 64, and 128.}
    \label{tab:caliper-wm}
          \begin{subtable}[t]{\linewidth}                                                          \centering                                                                               \caption{\textbf{Benchmark:} ParEval OpenMP Implementations using \texttt{GPT-4.1-mini} with Caliper Measurements}                                          \label{tab:caliper-wm-gpt41mini}                                                         \renewcommand{\arraystretch}{1.12}                                                       \setlength{\tabcolsep}{3.5pt}                                                            \begin{tabular}{@{}l|cc|cc|cc|cc|cc@{}}                                                  \toprule                                                                                 & \multicolumn{2}{c}{4 threads} & \multicolumn{2}{c}{16 threads} & \multicolumn{2}{c}{64 threads} & \multicolumn{2}{c}{128 threads} & \multicolumn{2}{|c}{Average} \\             \cmidrule(lr){2-3} \cmidrule(lr){4-5} \cmidrule(lr){6-7} \cmidrule(lr){8-9} \cmidrule(lr){10-11}                                                                                  \textbf{World Model} & Base & +SFT & Base & +SFT & Base & +SFT & Base & +SFT & Base & +SFT \\                                                                                  \midrule   
                    \texttt{Qwen-2.5-32B} & 42.7\% & \textbf{60.6\%} & 45.1\% & \textbf{60.8\%} & 41.7\% & \textbf{59.7\%} & 42.7\% & \textbf{57.8\%} & \cellcolor{gray!20} 43.0\% & \cellcolor{gray!20} \textbf{59.7\%} \\  
          \texttt{Qwen-2.5-7B}  & 46.8\% & \textbf{58.4\%} & 48.6\% & \textbf{56.9\%} & 45.7\% & \textbf{54.5\%} & 46.4\% & \textbf{53.5\%} & \cellcolor{gray!20} 46.9\% & \cellcolor{gray!20} \textbf{55.8\%} \\                      \texttt{Llama-3.1-8B} & 51.4\% & \textbf{60.2\%} & 50.2\% & \textbf{59.4\%} & 47.8\% & \textbf{57.1\%} & 47.7\% & \textbf{57.9\%} & \cellcolor{gray!20} 49.3\% & \cellcolor{gray!20} \textbf{58.6\%} \\ \midrule
          \textbf{Average} & \cellcolor{gray!20} 47.0\% & \cellcolor{gray!20} \textbf{59.7\%} & \cellcolor{gray!20} 48.0\% & \cellcolor{gray!20} \textbf{59.0\%} & \cellcolor{gray!20} 45.1\% & \cellcolor{gray!20} \textbf{57.1\%} & \cellcolor{gray!20} 45.6\% & \cellcolor{gray!20} \textbf{56.4\%} & \cellcolor{gray!20} 46.4\% & \cellcolor{gray!20} \textbf{58.0\%} \\\bottomrule                                                                              \end{tabular}                                                                     \end{subtable}  
              \vspace{0.0em}
              
    \begin{subtable}[t]{\linewidth}
        \centering
        \caption{\textbf{Benchmark:} ParEval OpenMP Implementations using \texttt{GPT-5.1} with Caliper Measurements}
        \label{tab:caliper-wm-gpt51}
        \renewcommand{\arraystretch}{1.12}
        \setlength{\tabcolsep}{3.5pt}
        \begin{tabular}{@{}l|cc|cc|cc|cc|cc@{}}
        \toprule
        & \multicolumn{2}{c}{4 threads} & \multicolumn{2}{c}{16 threads} & \multicolumn{2}{c}{64 threads} & \multicolumn{2}{c}{128 threads} & \multicolumn{2}{|c}{Average} \\
        \cmidrule(lr){2-3} \cmidrule(lr){4-5} \cmidrule(lr){6-7} \cmidrule(lr){8-9} \cmidrule(lr){10-11}
        \textbf{World Model} & Base & +SFT & Base & +SFT & Base & +SFT & Base & +SFT & Base & +SFT \\
        \midrule    
          \texttt{Qwen-2.5-32B} & 48.2\% & \textbf{50.5\%} & 41.2\% & \textbf{56.0\%} & 41.2\% & \textbf{56.1\%} & 41.2\% & \textbf{56.6\%} & \cellcolor{gray!20} 43.0\% & \cellcolor{gray!20} \textbf{54.8\%} \\       
        \texttt{Qwen-2.5-7B}  & \textbf{55.5\%} & 52.5\% & 47.9\% & \textbf{52.5\%} & 46.6\% & \textbf{49.1\%} & 44.6\% & \textbf{51.5\%} & \cellcolor{gray!20} 48.6\% & \cellcolor{gray!20} \textbf{51.4\%} \\
        \texttt{Llama-3.1-8B} & 48.5\% & \textbf{51.8\%} & 48.1\% & \textbf{57.9\%} & 44.2\% & \textbf{54.0\%} & 45.2\% & \textbf{55.4\%} & \cellcolor{gray!20} 46.5\% & \cellcolor{gray!20} \textbf{54.8\%} \\ \midrule
        \textbf{Average} & \cellcolor{gray!20} 50.7\% & \cellcolor{gray!20} \textbf{51.6\%} & \cellcolor{gray!20} 45.7\% & \cellcolor{gray!20} \textbf{55.5\%} & \cellcolor{gray!20} 44.0\% & \cellcolor{gray!20} \textbf{53.1\%} & \cellcolor{gray!20} 43.7\% & \cellcolor{gray!20} \textbf{54.5\%} & \cellcolor{gray!20} 46.0\% & \cellcolor{gray!20} \textbf{53.7\%} \\
        \bottomrule
        \end{tabular}
    \end{subtable}
\end{table}

\section{Conclusion}

We introduced Parallel-Code World Models (PCWMs), reasoning language models that predict tool outcomes directly from parallel source code. To train PCWMs, we developed a scalable data-collection pipeline that generates diverse parallel programming problems, candidate OpenMP implementations, tool executions, and chains of thought to causally connect source code to the observed outcomes. For ThreadSanitizer-based race analysis, we showed that reasoning supervision is important for effective world-model learning. We further showed that the resulting PCWMs provide useful feedback to downstream race-fixing agents, improving bug-fixing performance, showing potential to mitigate the challenges of external tool calling. Also, for Caliper-based performance profiling, fine-tuning led to improvements over the base models. 
Future work includes expanding PCWMs beyond OpenMP and beyond the two tools studied here. Another promising direction is to find means of better exploration for problem and candidate code generation to difficult and complicated situations pertaining to world modeling.


\section*{Acknowledgments}
This material is based upon work supported by the U.S. Department of Energy, Office of Science, Office of Advanced Scientific Computing Research, through solicitation DE-FOA-0003264, ``Advancements in Artificial Intelligence for Science," under Award Number DE-SC0025598.
This work was performed under the auspices of the U.S. Department of Energy by Lawrence Livermore National Laboratory (LLNL) under Contract DE-AC52-07NA27344 (LLNL-CONF-2017735). This research used resources of the Oak Ridge Leadership Computing Facility, which is a U.S. Department of Energy, Office of Science User Facility supported under Contract DE-AC05-00OR22725.


\bibliography{colm2026_conference}
\bibliographystyle{colm2026_conference}

\clearpage
\appendix
\section{Prompt Templates for Data Synthesis}
\label{app:prompts}

\subsection{Problem Variant Generation}

This prompt is used to generate diverse problem variants from seed problems. Each seed problem is perturbed to create 20 related but distinct parallel programming problems. We use the model \texttt{GPT-4o} for this and generate problem variants from 80 seed problems.

\begin{lstlisting}[style=promptstyle, caption={Problem Variant Generation Prompt Template}]
<problem>
{problem}
</problem>

# Instructions
For the following problem, propose {num_variants} parallel programming problems that are closely related but distinct from the above problem (a small perturbation).

In your perturbed problem variants, you can slightly change one or more of the following about the problem: (1) the input data type and data structure (e.g., arrays can become matrices, integers can become floats, etc.), (2) the algorithmic details (e.g., slightly different operations, slightly different control flow, etc.), (3) the problem statement, or (4) the size or dimensionality of the inputs.

# Formatting Instructions
Your response must only contain {num_variants} problem descriptions inside numbered tags as follows:

<variant_1>
{PERTURBED PROBLEM STATEMENT 1 GOES HERE.}
</variant_1>

<variant_2>
{PERTURBED PROBLEM STATEMENT 2 GOES HERE.}
</variant_2>

...

<variant_{num_variants}>
{PERTURBED PROBLEM STATEMENT {num_variants} GOES HERE.}
</variant_{num_variants}>

Do NOT include any additional text or explanations.
\end{lstlisting}


\subsection{Task-Specific Harness Generation}

This prompt is used to generate problem-specific harness code that will provide a reference serial implementation as well as code that samples test cases to drive the serial as well as parallel implementations. We use \texttt{GPT-4o} for this.

\begin{lstlisting}[style=promptstyle, caption={Reference Code Generation Prompt Template}]
<Makefile>
```
{Makefile}
```
</Makefile>

<harness.cc>
```cpp
{harness_cc}
```
</harness.cc>

<reference.cc>
```cpp
{reference_prompt_cc}
```
</reference.cc>

# Instructions
For the following problem, complete the `reference.cc` code by following the coding instructions given in-line:

<problem>
{problem}
</problem>

# Formatting Instructions
Your response must only contain a Markdown code block as follows:

```cpp
{THE COMPLETED `reference.cc` CODE GOES HERE.}
```

Do NOT include any additional code or explanations.
\end{lstlisting}


\subsection{Parallelization Generation With Differing Efficiencies}

This prompt is used to generate a parallelized version of the serial reference implementation. By providing the serial reference implementation, this LLM has clarity on the exact arguments and return type that the problem-specific harness expects. In addition, we ask the LLM in this step to generate diverse parallelized implementations with possible parallelization inefficiencies to provide more diversity and improve exploration quality. For this step, we use \texttt{Llama-3.3-70B-Instruct}, \texttt{Llama-3.1-8B-Instruct}, \texttt{Gemma-3-27B}, and \texttt{phi-4}. We use a temperature of 0.0 and ask for 4 diverse implementations within the same response.

\begin{lstlisting}[style=promptstyle, caption={Inefficient Parallelization Generation Prompt Template}]
<Makefile>
```
{Makefile}
```
</Makefile>

<harness.cc>
```cpp
{harness_cc}
```
</harness.cc>

<reference.cc>
```cpp
{reference_cc}
```
</reference.cc>

# Instructions
For the following problem, write {K} implementations for the required dependency file `generated.cc`.

<problem>
{problem}
</problem>

Keep in mind for each of the {K} implementations:
    1. Your `generated.cc` must include a definition of the `generated` function. Do not forget to `#include` anything that is necessary for the `generated` function.
    2. Your `generated.cc` needs to be compatible as a dependency in the compilation process provided in the given Makefile.
    3. Your implementation of the `generated` function must use a function signature identical to the one declared in `reference.cc`.
    4. Your implementation of the `generated` function must use OpenMP.
    5. Your implementation of the `generated` function must have some parallelization inefficiency, e.g., synchronization overhead, load imbalance, thread management overhead, inefficient scheduling, limited parallelism, etc.
    6. Do NOT write any code comments whatsoever within your generated code.
    7. Each of the {K} implementations must be distinct from one another in terms of the OpenMP directive(s) or the parallelization strategy used in order to exhibit distinct inefficiency scenarios.

# Formatting Instructions
Your response must only contain {K} Markdown code blocks inside numbered tags as follows:

<implementation_1>
```cpp
{THE COMPLETE `generated.cc` CODE GOES HERE. THERE MUST BE NO CODE COMMENTS.}
```
</implementation_1>

<implementation_2>
```cpp
{THE COMPLETE `generated.cc` CODE GOES HERE. THERE MUST BE NO CODE COMMENTS.}
```
</implementation_2>

...

<implementation_K>
```cpp
{THE COMPLETE `generated.cc` CODE GOES HERE. THERE MUST BE NO CODE COMMENTS.}
```
</implementation_K>

Do NOT include any additional code or explanations.
\end{lstlisting}


\subsection{Parallelization with Possible Race Conditions}

This prompt is used to generate a parallelized version of the serial reference implementation. By providing the serial reference implementation, this LLM has clarity on the exact arguments and return type that the problem-specific harness expects. In addition, we ask the LLM in this step to generate diverse parallelized implementations and produce diverse scenarios of race conditions to provide more diversity and improve exploration quality. For this step, we use \texttt{Llama-3.3-70B-Instruct}, \texttt{Llama-3.1-8B-Instruct}, \texttt{Gemma-3-27B}, and \texttt{phi-4}. We use a temperature of 0.0 and ask for 4 diverse implementations within the same response.

\begin{lstlisting}[style=promptstyle, caption={Racy Parallelization Generation Prompt Template}]
<Makefile>
```
{Makefile}
```
</Makefile>

<harness.cc>
```cpp
{harness_cc}
```
</harness.cc>

<reference.cc>
```cpp
{reference_cc}
```
</reference.cc>

# Instructions
For the following problem, write {K} implementations for the required dependency file `generated.cc`.

<problem>
{problem}
</problem>

Keep in mind for each of the {K} implementations:
    1. Your `generated.cc` must include a definition of the `generated` function. Do not forget to `#include` anything that is necessary for the `generated` function.
    2. Your `generated.cc` needs to be compatible as a dependency in the compilation process provided in the given Makefile.
    3. Your implementation of the `generated` function must use a function signature identical to the one declared in `reference.cc`.
    4. Your implementation of the `generated` function must use OpenMP.
    5. Your implementation of the `generated` function must have a data race.
    6. Do NOT write any code comments whatsoever within your generated code.
    7. Each of the {K} implementations must be distinct from one another in terms of the OpenMP directive(s) or the parallelization strategy used in order to exhibit distinct data race scenarios.

# Formatting Instructions
Your response must only contain {K} Markdown code blocks inside numbered tags as follows:

<implementation_1>
```cpp
{THE COMPLETE `generated.cc` CODE WITH DATA RACE GOES HERE. THERE MUST BE NO CODE COMMENTS.}
```
</implementation_1>

<implementation_2>
```cpp
{THE COMPLETE `generated.cc` CODE WITH DATA RACE GOES HERE. THERE MUST BE NO CODE COMMENTS.}
```
</implementation_2>

...

<implementation_K>
```cpp
{THE COMPLETE `generated.cc` CODE WITH DATA RACE GOES HERE. THERE MUST BE NO CODE COMMENTS.}
```
</implementation_K>

Do NOT include any additional code or explanations.
\end{lstlisting}

\clearpage

\subsection{Chain-of-Thought for Data Race Detection}

This prompt is used to generate Chain-of-Thought for the ThreadSanitizer, mimicking world model training. We specifically highlight in the prompt to in-fill the chain-of-thought rather than discussing it as a foregone conclusion to analyze. We use \texttt{QwQ-32B} for this prompt with temperature 0.999, \texttt{top\_p}=0.95, and a maximum tokens of 32768.

\begin{lstlisting}[style=promptstyle, caption={Data Race Detection CoT Prompt Template}]
# Given {execution_model} Code
```cpp
{generated_cc}
```

# Instructions
Your job is to think and predict data races that would be caught by ThreadSanitizer in the above {execution_model} code.

## Step 1: Think
- Inside the tags <think> and </think>, in-fill a long chain of thought for the given {execution_model} code line by line.
- In this chain of thought, go over each code line, write that code line, what the code line does and a verbose logical explanation why that line would or wouldn't cause lead ThreadSanitizer to catch a data race. Thoroughly explore every important or non-trivial code line or code block by engaging in a comprehensive cycle of analysis, summarizing, exploration, reassessment, reflection, back-tracking, and iteration to develop well-considered thinking process about the detection of data race.
- Inside <think> and </think>, NEVER mention or talk about you having prior knowledge of data race(s) present or prior knowledge of races thrown by ThreadSanitizer. You are only in-filling the thought for the given code.
- The conclusion of the in-filled chain of thought, however, must lead to the correct answer i.e., the ThreadSanitizer's race detections as provided below. Therefore, you will engage in a comprehensive cycle of analysis, summarizing, exploration, reassessment, reflection, back-tracking, and iteration until the correct conclusion is reached. Remember, NEVER mention or talk about you having prior knowledge of data race(s) present or prior knowledge of races thrown by ThreadSanitizer.

## Step 2: Answer
After writing the thought inside the tags <think> and </think>, write verbatim (remember verbatim) the following answer inside the tags <answer> and </answer>:
<answer>
\nHere is a list of data races that ThreadSanitizer will catch:\n\n

{race_outcome}

</answer>

# Formatting Instructions
In summary, your overall response will be:

<think>
{IN-FILLED CHAIN OF THOUGHT GOES HERE LEADING TO CORRECT CONCLUSION.}
</think>
<answer>
\nHere is a list of data races that ThreadSanitizer will catch:\n\n

{race_outcome}

</answer>
\end{lstlisting}


\subsection{Chain-of-Thought for Performance Prediction}

This prompt is used to generate Chain-of-Thought for the Caliper mimicking world model training.  The model is given two code implementations and their Caliper profiling results, and must generate reasoning that appears to predict the performance differences. We specifically highlight in the prompt to in-fill the chain-of-thought rather than discussing it as a foregone conclusion to analyze. We use \texttt{QwQ-32B} for this prompt with temperature 0.999, \texttt{top\_p}=0.95, and a maximum tokens of 32768.

\begin{lstlisting}[style=promptstyle, caption={Performance Prediction CoT Prompt Template}]
# Code A
```cpp
{generated_cc_a}
```

# Code B
```cpp
{generated_cc_b}
```

# Broad Definition of Caliper's Work Percentage
The percentage of total OpenMP thread time spent doing useful application work, as opposed to OpenMP runtime overhead (e.g., barriers, synchronization, idle time, etc.).

# Instructions
Code A and Code B are two alternative implementations of the same functionality. Your job is to think and predict work percentages of the given Code A and Code B as would be measured by Caliper by identifying analogous parallel region pairs from Code A and Code B and focusing on difference in work percentages between regions within a pair.

## Step 1: Think
- Inside the tags <think> and </think>, in-fill a long chain of thought analyzing Code A and Code B.
- In this chain of thought, identify **pairs of analogous parallel region** from Code A and Code B.
- Go over each pair of analogous parallel regions, analyzing the two parallel regions together side by side in terms of their differences.
- Thoroughly explore these analogous parallel region pairs engaging in a comprehensive cycle of comparison, analysis, summarizing, exploration, reassessment, reflection, back-tracking, and iteration to develop well-considered thinking process about work percentage and how the work percentages measured by Caliper would differ between the two parallel regions.
- Inside <think> and </think>, NEVER mention or talk about you having prior knowledge of Caliper measurements. You are **in-filling** or **smoothing** the chain of thought analyzing codes A and B -- NOT post-hoc explaining of Caliper measurements.
- The conclusion of the in-filled chain of thought, however, must lead to the correct answer i.e., the correct relative difference in Caliper measurements between analogous parallel regions from codes A and B. Therefore, you will engage in a comprehensive cycle of analysis, summarizing, exploration, reassessment, reflection, back-tracking, and iteration until the correct conclusion is reached. Remember, NEVER mention or talk about you having prior knowledge of Caliper measurements.

## Step 2: Answer
After writing the thought inside the tags <think> and </think>, write verbatim (remember **verbatim**) the following answer inside the tags <answer> and </answer>:
<answer>

## Caliper Measurements for Code A
<measurements>
{measurements_a}
</measurements>

## Caliper Measurements for Code B
<measurements>
{measurements_b}
</measurements>

</answer>

# Formatting Instructions
In summary, your overall response will be:

<think>
{IN-FILLED CHAIN OF THOUGHT GOES HERE LEADING TO CORRECT ANSWER.}
</think>
<answer>

## Caliper Measurements for Code A
<measurements>
{measurements_a}
</measurements>

## Caliper Measurements for Code B
<measurements>
{measurements_b}
</measurements>

</answer>
\end{lstlisting}

\subsection{Caliper Instrumentation Procedure}
We automatically instrument the generated OpenMP code with Caliper profiling markers using a four-step process:
(1) An LLM (\texttt{GPT-4o}) takes the candidate code with line numbers and identifies all OpenMP parallel regions by returning their start and end line numbers as a JSON array.
(2) For each identified region spanning lines $[s, e]$, we insert \texttt{CALI\_MARK\_BEGIN("region\_s\_e")} before line $s$ and \texttt{CALI\_MARK\_END("region\_s\_e")} after line $e$, preserving code indentation.
(3) Markers are inserted in reverse order (bottom to top) to maintain original line number references, with proper handling of nested regions by processing larger spans first.
(4) The Caliper header \texttt{\#include <caliper/cali.h>} is added at the top of the file, yielding instrumented code ready for profiling with thread counts 4, 16, 64, and 128. We leverage Hatchet \citep{hatchet} to read the Caliper reports.

\begin{lstlisting}[style=promptstyle, caption={Caliper Instrumentation Region Identification Prompt Template (GPT-4o)}]
# Given OpenMP Code with Line Numbers
```cpp
{generated_cc_with_line_no}
```


# Instructions
Identify all OpenMP parallel regions. Return the identified regions as a JSON array with each element in the array containing integer fields called 'start' and 'end'.

Specifically,
- Focus on identifying all OpenMP parallel regions in the given code (start = #pragma line, end = closing brace of that parallel region).
- Regions may be nested as long as each region captures a distinct #pragma and its associated parallel region.
- Use the original line numbers from the given code above in your returned JSON array to demarcate the regions.


# Formatting Instructions
Your response must ONLY contain a valid Markdown JSON block containing a JSON array like:
```json
[
  {{'start': integer, 'end': integer}},
  {{'start': integer, 'end': integer}},
  ...
]
```
Do NOT include any additional code or explanations, just the valid Markdown JSON block.
\end{lstlisting}

\section{Training Hyperparameters}

Table~\ref{tab:hyperparameters} summarizes the training hyperparameters used.

\begin{table}[h]
\centering
\caption{\textbf{World Model Fine-Tuning Hyperparameters.} Training configuration used to adapt the base models into reasoning world models.}
\label{tab:hyperparameters}
\begin{tabular}{@{}p{0.31\linewidth}c@{}}
\toprule
\textbf{Hyperparameter} & \textbf{Setting} \\
\midrule
Sequence Length & 16,384 tokens \\
Precision & \texttt{bfloat16} \\
FSDP Sharding Strategy & Full-shard \\
Gradient Checkpointing & Enabled \\
Number of Epochs & 1 \\
Effective Batch Size & 32 \\
Learning Rate & $1\times10^{-5}$ \\
Learning Rate Scheduler & Cosine \\
Warmup Ratio & 0.05 \\
Weight Decay & $1\times10^{-4}$ \\
AdamW $\beta_1$ & 0.9 \\
AdamW $\beta_2$ & 0.95 \\
Validation Split & 5\% \\
Completion-Only Loss & Enabled \\
\bottomrule
\end{tabular}
\end{table}

\clearpage
\section{Additional Experimental Results}

\subsection{Race-Fixing using Commercial Actor Models on DataRaceBench}

\begin{table}[h]
\centering
\caption{\textbf{Race-Fixing using Commercial Actor Models on DataRaceBench.} We task an LLM to act as a race-fixing agent to fix race conditions in OpenMP codes in DataRaceBench. The race-fixing agent receives race analysis and detection feedback from the world model. The race-free percentages of the fixed codes are reported as measured using ThreadSanitizer.}
\label{tab:race-fixing-commercial}
\vspace{0.5em}
\setlength{\tabcolsep}{4pt}
\begin{tabular}{@{}l|c|c|cc|cc|cc@{}}
\toprule
 & \multicolumn{8}{c}{Feedback Source} \\ \cmidrule{2-9}
 & \multirow{2}{*}{Oracle} & \multirow{2}{*}{Self} & \multicolumn{6}{c}{World Model} \\
\cmidrule(l){4-9}
Actor Model &  &  & \multicolumn{2}{c|}{\texttt{Qwen-2.5-32B}} & \multicolumn{2}{c|}{\texttt{Qwen-2.5-14B}} & \multicolumn{2}{c}{\texttt{Qwen-2.5-7B}} \\
\cmidrule(lr){4-5}\cmidrule(lr){6-7}\cmidrule(lr){8-9}
 &  &  & Base & +SFT & Base & +SFT & Base & +SFT \\
\midrule
\texttt{claude-haiku-3} & 72.2\% & 71.2\% & 71.1\% & \textbf{73.0\%} & \textbf{72.5\%} & 69.9\% & 69.3\% & \textbf{70.6\%} \\
\texttt{gpt-4.1-mini}   & 83.2\% & 83.1\% & 77.9\% & \textbf{82.6\%} & 79.9\% & \textbf{81.1\%} & 77.7\% & \textbf{79.4\%} \\
\midrule
\rowcolor{gray!10} Average & 77.7\% & 77.2\% & 74.5\% & \textbf{77.8\%} & \textbf{76.2\%} & 75.5\% & 73.5\% & \textbf{75.0\%} \\
\bottomrule
\end{tabular}
\end{table}

\section{Additional Related Work}

\textbf{CoT Synthesis.} CoT synthesis and reasoning distillation are increasingly used to turn base LLMs into reasoning-capable models, as illustrated by OpenThoughts others \citep{guha2025openthoughts,bespoke2025stratos,novasky2025skyt1, muennighoff2025s1}. One of the techniques for CoT synthesis involves \emph{reverse} rationale generation conditioned on the correct answer when initial reasoning fails \citep{zelikman2022star, zelikman2024quietstar, bhagavatula2020abductive, li2025evipath}. Yet, these methods do not target CoTs for tool-outcomes, unlike ours.

\end{document}